# α-Al$_2$O$_3$ sapphire and rubies deformed by dual basal slip at intermediate temperatures (900°C-1300°C). II – Dissociation and stacking faults

**M. Castillo Rodríguez\*, J. Castaing\*\*, A. Muñoz\*, P. Veyssière\*\*\* and A. Domínguez Rodríguez\***

\*Departamento Física de la Materia Condensada. Universidad de Sevilla, anmube@us.es. Apdo. 1065, 41080 Seville (Spain)

\*\*C2RMF, CNRS UMR 171, jacques.castaing@culture.fr. Palais du Louvre, 14 quai François Mitterrand 75001 Paris (France)

\*\*\*LEM, CNRS UMR 104, ONERA, BP 72, 92322 CHATILLON Cedex (France)

**Abstract :**

Sapphire and rubies (undoped and Cr-doped α-Al$_2$O$_3$ single crystals) have been deformed in compression at temperatures lower than those previously used in studies of dislocations in the basal slip plane (see part I). Above 1400°C, several features associating stacking faults out of the basal planes and partial dislocations (dissociation, faulted dipoles) have been observed in previous transmission electron microscope investigations. The formation of these features involves climb controlled by atomic diffusion. Properties of climb dissociated dislocations are discussed in relation with dislocation dynamics. TEM examination of dislocation structures at lower deformation temperatures (1000-1100°C) shows that similar features are formed but that they often imply cross-slip. A new mechanism for the formation of faulted dipole by glide is presented and an explanation for the 30° Peierls valley orientation is proposed. The presence of chromium has a small influence on stacking fault energies on planes perpendicular to the basal plane.

## 1. Introduction

In the first part of this work (part I), we have investigated the dislocation microstructures introduced in sapphire and ruby (undoped and Cr-doped α-Al$_2$O$_3$ single crystal) deformed under a specific load orientation, dubbed *aa*. The deformation temperature at which the behaviour of basal dislocations can be explored in samples void of deformation twins is lowered to near the brittle to ductile transition (900-1000°C) [1]. In decreasing the deformation temperature, we have found situations where dislocation climb is no longer prominent and where glide properties could be unambiguously investigated. The *aa* orientation thus provides a unique opportunity to further investigate the temperature dependence of dislocation dissociation in relation with the macroscopic behaviour in a domain where climb activity is susbstantially reduced, if not eliminated. So far, It is ascertained that above 1400°C in sapphire basal dislocations dissociate by climb into two partials bordering a stacking fault SF according to [2-6]

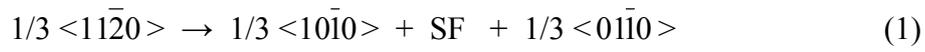

$$1/3 <11\bar{2}0> \rightarrow 1/3 <10\bar{1}0> + SF + 1/3 <01\bar{1}0> \qquad (1)$$

It is generally agreed that in the course of a deformation test at elevated temperature, whenever a non-screw dislocation is arrested at some obstacle, climb-dissociation then nucleates and proceeds, forming a sessile configuration [4, 7]. The hindering role of climb-dissociation decreases with test temperature. Further to dissociated dislocations, several faulted configurations such as faulted dipoles and faulted loops have been reported and their formation mechanisms discussed in terms of the non-conservative motion of partial dislocations [4, 6, 8, 9]. As for dislocation splitting, there no indication from deformation debris that they could exist as individual partials whereas the collective activity of 1/3 $<01\bar{1}0>$ partials is attested by basal twinning [10]. This is confirmed by the absence of evidence from transmission electron microscope (TEM) observations and numerical simulations [2, 4, 11] that 1/3<1120> basal dislocations should be dissociated in their glide plane.

In this context, the present paper focuses on the fine structure of dislocations and their debris in sapphire and rubies deformed by basal slip to the lower yield point, down to 900-1000°C, with special attention paid to the influence of the Cr dopant.

## 2. Results and discussion

Details on the experimental procedure can be found in part I. It is worth emphasizing that further to the foils sliced parallel to the basal plane used in Part I, other foils with normals far away from [0001], hence more suitable to analyze climb-dissociated dislocations [4, 6], were utilized.

2.1. Dissociation distance and habit plane.

Figure 1 shows the dissociation of basal dislocations (reaction (1)) in sapphire and ruby doped with two different concentrations of $Cr^{3+}$ (725 and 9540 mol ppm), after deformation 1300°C to a permanent strain of about 2%. <span style="color:red">The dislocation undulation noted in</span> Figure 1<span style="color:red">(a) and (b) will be discussed in section 2.4.</span> In the [$22\bar{4}1$] foil orientation, climb dissociation is observed along the long segments parallel to [$1\bar{1}00$] which are either edge in character with Burgers vector $\mathbf{b_1}$ = 1/3[$11\bar{2}0$] or 30° dislocations with Burgers vector $\mathbf{b_2}$ =1/3 [$\bar{2}110$] (Part I). We have in addition analyzed foils whose [$\bar{5}051$] normal, located between -$\mathbf{b_1}$ and $\mathbf{b_2}$, enables the characterization of climb dissociation of 60º dislocations with either one of these Burgers vectors. Within the experimental uncertainty, dislocations are dissociated to about the same amount as the edge segments, i.e. 6 to 8 nm (see the Appendix for the dependence of the driving forces for climb dissociation on dislocations character), consistent with previous weak-beam TEM estimates performed on climb-dissociated edge dislocation in *a*-oriented specimens [2, 4, 5, 8]. It is thus clear that the stacking fault energy does not change substantially between {$10\bar{1}0$} and {$\bar{1}\,\bar{1}20$} planes, in accord with shell-model computations [9]. We have in addition found that in ruby deformed at 1100 ºC and containing 725 ppm Cr (close to sapphire), dislocations exhibit climb-dissociated configurations similar to those observed in the same material after deformation at 1300°C (Figure 1(b)). In all cases, the dissociation width amounts to 6 to 8 nm (Figure 1). It is worth emphasizing that in principle 30º basal dislocations can dissociate in a $\{\bar{1}\bar{1}20\}$ prism plane where one of its partials, and one only, can cross-slip thus fully immobilizing this segment [3]. This property provides a simple interpretation of the preference of 30° dislocations after deformation at low temperatures in ruby and sapphire (part I).

To further characterize dissociation properties in sapphire and the influence of doping on climb dissociation, we have conducted a dedicated determination of the stacking fault habit plane and of the dissociation width in this plane. This was achieved through a stereographic analysis of dislocations introduced by deformation at 1300°C. The distance $d$ between the images of the two partial dislocations is shown as a function of the angle $\theta$ between the electron beam and the [0001] direction in *Table 1* (Figure 2). The angles $\alpha$, $\beta$, and $\theta$ are related by :

$$\cos(\alpha + \beta) = \sin\theta \qquad (2)$$

where $\beta$ is a measure of the deviation of the dissociation plane from the plane perpendicular to the basal plane, corresponding to pure climb dissociation. One has:

$$\cos\left(\cos^{-1}\left(\frac{d}{h}\right) + \beta\right) = sin\theta \qquad (3)$$

The dissociation distance $h$ and the stacking fault habit plane can be determined from this formula and the various ($\theta_i$, $d_i$) pairs measured experimentally (*table 2*). Leaving apart the influence of the undulation (see section 2.4), a large fraction of the uncertainty on $\beta$ and $h$ arise from two main factors. On the one hand, the dislocation signal originates from a region in the core vicinity adequately distorted to satisfy the Bragg orientation, off the actual dislocation line position [10] and on the other, the partials which have non-parallel Burgers vectors, are viewed under distinct **g.b** conditions. Their images are thus shifted differently with respect to the actual positions of the partials. This latter property is also responsible for the so-called outside/inside dipolar effect whereby the images of the partials are either pulled apart or brought closer depending on the sign of the product $s_g$ **g.b**. The effect, which has been analysed in detail by Lagerlöf et al. [8, 11], does not influence the determination of the habit plane provided the sign of $s_g$ **g.b** is unchanged during the tilting experiment. This is, however, no longer true for the dissociation distance. The image simulations suggest that the actual distance between partial is less by 20% than the distance between images for "outside" contrast [8, 11], smaller than the dispersion of experimental data (table 1). A more detailed study is therefore useless, in particular since we have not attempted to evaluate a stacking fault energy, restricting ourselves to an overall appreciation of trends from non-corrected measured distances.

One can see from *table 2* that $Cr^{3+}$ impurities do not modify the stacking fault habit plane which remains nearly perpendicular to the (0001) plane in all cases ($\beta \approx 0$). Dissociation thus occurs essentially by pure climb as this had been previously reported for sapphire [2, 4, 5, 8]. As for dissociation, *table 2* shows that the mean value of $h$ amounts to about 6.5 nm in sapphire, a value which lies between the separation of 8 nm determined by Lagerlof *et al* [8] under weak beam and the separations of 4.7 nm [5] and 3.5 nm [12] under HRTEM. In spite of the above-mentioned uncertainty, it can be reasonably concluded that the separation distance $h$ increases typically from 6.5 nm for undoped or lightly doped sapphire, to 8.5 nm in samples doped with the highest chromium concentration, indicating that the stacking fault energy is moderately but perceivably changed by addition of $Cr^{3+}$ to sapphire. Down to 1100°C, temperature has little influence, if any, on dissociation out of the basal plane.

## 2.2. Climb dissociation and plastic properties

The question arises as to which extent the stacking fault energy reduction by Cr additions could affect mechanical properties in the range of temperatures where dissociation by climb operates. When climb-dissociated, dislocations are unable to drag a stacking fault (the backstress given by $3^{1/2}\gamma_{bas}/b$ where $\gamma_{bas}$ is the stacking fault energy in the basal plane, amounts to about 7.3 GPa for $\gamma_{bas} = 2$ J.m$^{-2}$) and they are therefore sessile in basal slip unless they can recombine forming a glissile core [7]. Because of the mechanical work needed for recombination, it is sometimes argued that the larger the separation between partials, the larger the flow stress. This scenario was, however, not confirmed experimentally by Castillo *et al.* [13] who observed but a small change in CRSS with Cr concentrations whose origin is consistent with elastic interactions between $1/3<11\bar{2}0>$ dislocations and impurities. Moreover, the CRSS dependence on temperature is smooth and monotonous for sapphire [1, 3, 14] as well as for rubies [13], again suggesting that there is no influence of climb dissociation on plastic deformation, a conclusion further supported by the quasi absence of yield points after static ageing [1].

It is then interesting to understand why the formation of sessile dislocations in overwhelming proportions should have little effect, if any, on the mechanical properties of sapphire and rubies. In order to appreciate the stresses involved in the transformation from a climb-dissociated to a glissile configuration, one may consider the stress experienced by one partial on its way toward recombination. When, for instance, this partial is located at 5$b$ from its companion, the interaction stress is actually of the order of several GPa, clearly in excess of the CRSS. It is therefore pointless to assume that the recombination of a climb-dissociated dislocation could be assisted by the external stress, let alone the necessary matter transport and rearrangement between partials. In this respect, we believe that the conclusions drawn by Shibata *et al.* [12] on the core structure of gliding basal dislocations in sapphire are largely unfounded. This is mainly because the transformation of a climb-dissociated configuration into a compact core glissile in the basal plane, an essential ingredient in the Shibata *et al*'s reasoning, is regarded as a mere geometrical change. Should a mechanism for the direct transformation from a climb-dissociated into a compact gliding dislocation exist, this would involve dramatically high stresses together with significant diffusive reorganization and non linear distortions, while, in order to extrapolate the core structure of a gliding basal configuration directly from that of a climb-dissociated one, Shibata *et al*. [12] implicitly consider this transformation as spontaneous. The experimental evidence together with a recent computer simulation [15] would rather support a scenario where dislocations adopt a compact core in gliding in the basal plane [2, 13, 14] until they are arrested by interactions with their environment. At this stage they transform into more or less widely climb-dissociated dislocations, themselves intrinsically sessile. This latter dissociation mode can be thought of as occurring most profusely upon unloading at a rate dictated by atomic diffusion, that is, by the temperature of the test thus consistent with the observation that dislocations are almost systematically dissociated by climb in post mortem samples. By no means may this observation provides relevant indication on the actual structure of the cores of moving dislocations.

2.3. Constricted climb-dissociated dislocations

The apparent insensitivity of mechanical properties to climb dissociation might imply either that there is always a large amount of undissociated dislocations available for glide in sapphire or else that dissociated dislocations can be easily rendered mobile, for instance by taking advantage of constrictions (also dubbed nodes) along the stacking fault ribbon. Although the former mechanism is possible but difficult to prove, only the latter possibility is actually consistent with observations. The constrictions that are observed in places on climb-dissociated dislocations (e.g. C in Figure 1.c and perfect loop PL in Figure 4.a), previously noticed by Lagerlöf [11, 16]), deserve special attention for they cannot be understood as being formed once dislocations are fully dissociated by climb. This would indeed imply that the partials of a climb-dissociated dislocation are able to recombine at some special locations, a property which we have ruled out in the previous section on account of the extremely large driving stresses thus required. It is worth mentioning that they are true constrictions, not merely the crossing in projection of the images of two partial dislocations [16]. This can be understood by considering how climb dissociation should nucleate in practice. There is in fact no reason why in climb dissociating, one given partial should always be ascribed to climb "up" and its companion "down". Rather, climb dissociation is a succession of thermally-activated events where the partial that first climbs "up" is the partial that has first absorbed a vacancy emitted by its companion in climbing "down". It is reasonable to postulate that, in the vicinity of the nucleation site, this first mutual exchange of a point defect determines the way companion partials should continue to climb by mutual exchange of point defects. Suppose now that before the above climb-dissociated portion had time to expand sideways, a similar nucleation event occurs elsewhere on a still compact core portion of the dislocation under consideration. Then there is a finite probability for the partial that had climbed "down" at the other site, to climb "up" in this new location. The two segments are dissociated in opposite directions with respect to the basal slip plane. Climb dissociation subsequently proceeds to equilibrium separation, expanding from both nucleation sites until the two dissociated portions merge at which point the partial that had climbed up on one side must connect to the partial that has climbed down on the other side, and *vice versa*. In doing so the two partials must pass through a unique configuration at which they both belong to the basal plane. We believe that this is where the constriction is formed as a result of the high value assumed by the stacking fault energy calculated in this plane [17] itself supported by the lack of experimental evidence. It is noted that the constrictions in question might serve as nuclei for further core recombination in the basal plane forming glissile basal dislocation segments.

2.4. Zigzagged dissociated dislocations.

In accord with previous observations in sapphire deformed at 1400-1500°C [6, 8], we have very commonly observed that the dissociated dislocations are not straight but zigzagged. This is true in sapphire deformed at 1300°C (Figure 1*a*) as well as in 725ppm Cr-doped ruby deformed at 1100°C. The deviation from the mean line direction amounts to about ±30º, the zigzag periodicity is of approximately 50 nm and no constrictions are visible. This zigzag, whose amplitude in sapphire is about 10 nm, is noticeably less pronounced upon increasing chromium concentration (Figure 1). We have determined that the habit plane of the zigzagged dislocation does not deviate perceivably from the prism plane, and therefore that zigzagging proceeds essentially by pure climb.

We note that since the configuration is all coplanar in the prism plane, no explanation in terms of a possible anisotropy of the stacking fault energy may apply. By making use of the method developed in [18], we have checked that the elastic anisotropy of sapphire, which is weak in the prism plane too, is responsible for no line instabilities which would have given rise to dislocation faceting along favoured orientations. Another interpretation could rely on climb being locally enhanced by radiation damage. We, however, notice that whereas the zigzag amplitude is significantly lessened by addition of chromium, there is no indication that radiation damage might depend on the Cr content (Figure 1). Furthermore, the absorption of radiation damage by dissociated dislocations should give rise to local widenings instead of the observed rather regular zigzags, and there is no sign of a depletion of radiation damage in the vicinity of zigzagged dislocations.

One is left with an interpretation in terms of Peierls valleys for partial dislocations that would prefer $< 2\,\overline{1}\,\overline{1}\,1 >$ and $< 20\overline{2}1 >$ directions in $\{0\,\overline{1}\,10\}$ and $\{1\overline{2}10\}$ planes, respectively. A high chromium concentration would result in shallower minima thus smoothing out the segmentation.

## 2.5. Faulted dislocation dipoles and loops formed by climb

Although other loops or hairpin-like dipoles are not observed in the vicinity of the perfect elongated loop (a) shown in Figure 3, this loop is likely to have resulted from the diffusion-mediated pinching off of an elongated dipole (part I), a mechanism well-documented in sapphire [4, 6, 19, 20]. We have determined that its widened extremities (dipole separation about 0.1 μm) lie in a $(11\,\overline{2}\,0)$ prism plane, while the central zone is at about 55º from the basal plane, thus close to the dipole equilibrium configuration predicted by elasticity for an edge dipole (part I). The configuration of the loop extremities can be adopted only if they have transformed by climb [6] which is compatible here with deformation having been carried out at 1300°C. Loop (a) in Figure 3 is reminiscent of a previous observation of Phillips *et al.* [6] in *a*-orientated sapphire. Rather similar in shape at first sight, defects (a) and (b) (Figure 3) differ in that (a) is made of one continuous $1/3[11\,\overline{2}\,0]$ dislocation forming a loop whereas (b) is a dissociated dislocation loop (width about 60 nm) lying in a plane at 30° from the basal plane that is connected to a faulted dipole (separation about 20 nm) itself located in a plane perpendicular to the basal plane.

The (a) configuration of Figure 3 can be regarded as an illustration of an early step of the mechanism eventually yielding faulted dipoles and faulted loops [21], that is, before the two partials with opposite signs in the inner part have begun to mutually annihilate. The next step of the transformation is exemplified by loop (b) in Figure 3 with the narrow faulted dipole attached to a perfect loop. Faulted dipoles have been commonly observed in sapphire deformed in basal slip at temperatures higher than those investigated here [6, 21-23]. They can be regarded as the signature of deformation temperatures sufficiently high to promote rapid transformations by diffusion. It is indeed likely that the transformation takes place under a self-fed process where the point defects emitted during the approach of the inner dislocations migrate by pipe diffusion towards the dipole extremities forcing them to widen by climb. Faulted dipoles have been reported in previous studies [6, 21-23] and they are rather abundant in our samples too (see part I). The ratio of the number of faulted to unfaulted dipoles, FD/PD, increases from 0.7 to 1.5 when when the temperature is increased from 1000 ºC to 1300 ºC, in agreement with an extrapolation made from the literature [23].

## 2.6. Glissile faulting

In addition to the profuse dipolar defects such as (b) in Figure 3 where the faulted portion is narrower than the perfect dipole [6], we have observed configurations where the widths of faulted and unfaulted dipoles are almost the same (Figure 4(a)). Such a configuration is clearly inconsistent with the above-discussed model via diffusion-assisted dipole/loop width fluctuations and self-climb. It suggests a specific faulting process. In fact, the perfect basal dislocations of Figure 4(a) are 30º mixed in character ($[1\bar{1}00]$ line and $\mathbf{b} = 1/3\,[\bar{2}110]$). We have seen in § 2.1 that a 30° perfect dislocation always comprises a screw partial which can therefore cross-slip on the prism plane to combine with its dipolar partner dislocation leaving a faulted dipole in the prism plane, bordered by two attractive 60° partials with anti-parallel Burgers vectors. The segment separating the unfaulted and faulted parts is of course glissile in the dipole habit plane which coincides with the observed $\{11\bar{2}0\}$ prism plane slip, at variance from the cross-slip of perfect dislocations analyzed in § 3.2 of Part I.

The energy balance of this transformation can be deduced from equations 5-16 in [24] that gives the interaction energy of two parallel dislocations brought together from some distance $R_a$ to a distance R. Applied to a perfect screw dipole transforming into a faulted dipole bordered by two 30° partials (Figure 4(b)), these equations yield the following energy variation:

$$\Delta W/L = 2.55 \ \mu \ b^2/8\pi \ \ln(R/R_a) \ + \ \gamma R \qquad (4)$$

With $\mu = 157$ GPa, $b = 0.475 \ 10^{-9}$ m, $R_a = 100 \ R$, $\gamma = 0.2$ J/m² and $\nu = 0.25$, one finds that the faulted and unfaulted dipoles have the same energy for $R = 83$ nm. In other words, elasticity predicts that perfect and faulted dipoles should be energetically favoured for heights larger and smaller than 83 nm, respectively. This is in fairly good agreement with the separation of 50 nm reported for faulted dipoles [4, 6, 21, 25] and with the present TEM determination of ~~3~~20-50 nm (Figure 3 and Figure 4(a)). The driving stress for the faulting mechanism which amounts to 450 MPa for a separation of 25 nm, lies in the range of CRSS values measured for prism plane slip at 900-1000°C [3]. The applied stress may either help or hinder the phenomenon. Because of the hyperbolic dependence of the driving stress on the dipole separation, $R$, narrow dipoles ($R < 83$ ~~79~~ nm) should remain faulted irrespective of the applied stress whereas the nature of wide dipoles is strongly stress- (hence temperature-) dependent. The mechanism of Figure 4(a) observed at 1300°C should become more common below this temperature when diffusion ceases to operate. It may explain in particular the observation of faulted loops after deformation under hydrostatic pressure at 800°C [3].

As for pseudo dipoles, the above mechanism involving a reaction between two partials yields a similarly faulted configuration when the two pseudo-dipolar dislocations are both 30° in character. They indeed exhibit a pair of screw partials with antiparallel Burgers vectors which may therefore mutually annihilate by glide. The resulting faulted pseudo-dipole is comprised of the remaining two 60° partials which, at variance from the dipole case, exhibit non collinear Burgers vectors. This configuration has not been identified by TEM so far. If the two pseudo-dipolar dislocations are 30° and edge, only one partial out of four is screw and one can easily show that there is no driving force for cross-slip annihilation.

It is likely that a variant of the preceeding glide-mediated faulting mechanism has operated in the configuration shown in Figure 5 which consists in a kinked rectilinear dislocation whose kink exhibits a widely faulted section. This configuration was observed after a two-step deformation test where the first step consisted in 1.7% deformation at 1250°C, and in the second step, the sample was maintained under an applied stress of 451 MPa at 804 °C. We have determined that (i) the largest distance between partials is 76 nm, i.e. 10 times the equilibrium climb-dissociation width, (ii) the parent dislocation lies along $[1\bar{2}10]$, (iii) it is dissociated in two partials, $1/3[\bar{1}010]$ and $1/3[\bar{1}100]$, (iv) the $1/3[\bar{1}100]$ partial is bowed out out of the basal plane, (v) its companion is oriented along the $[\bar{1}100]$ direction and (vi) the fault habit plane $(11\bar{2}0)$ is perpendicular to the basal plane. This configuration is crystallographically compatible with the cross-slip dissociation mechanism of 30° dislocations in a prism plane mentioned in § 2.1. An explanation is that since the parent dislocation is stabilized by Peierls effects in the 60° orientation, its motion is achieved by superkinks which can locally adopt the 30° character allowing cross-slip of one partial. It is noted that at 750°C, the CRSS for prism plane and basal slips are comparable amounting to about 600 MPa [3]. For a stacking fault energy in prism planes, $\gamma_{prism}$, of the order of 0.2 mJ/m² [2, 4, 8], a shear stress of about 600 MPa (= $3^{1/2}\gamma_{prism}/b$) is enough to free a partial dislocation from its companion. This stress is of the order of magnitude of the stress applied at 804°C. The faulted configuration of Figure 5 bears similarities with the mechanism proposed by Lagerlof *et al.* [3] for the formation of faulted dipoles and loops at low temperatures and hydrostatic pressure, except that in this mechanism the curved partial is sessile and the step on the screw dislocation is a jog, not a kink.

**3. Conclusion**

TEM examination of dislocations after deformation at 1000-1300°C of sapphire and rubies reveals many features similar to those observed previously after deformation above 1400°C. Dissociation out of the slip plane is observed down to 1100°C, a temperature at which off-slip-plane configurations can be attained by glide in certain orientations. The later mechanism is essentially activated at low temperature at 30° dislocation where one partial is screw and can glide in a prism plane. This mechanism is consistent with the formation of faulted features in sapphire deformed at low temperature. The stacking fault energy in the prism plane is roughly the same in $\{11\overline{2}0\}$ and $\{1\overline{1}00\}$ planes. It decreases slightly in the presence of about 1% chromium in sapphire. Climb dissociation is largely irreversible but glide might be reactivated at constrictions that are formed randomly on climb-dissociated dislocations.

**Acknowledgements**

We are gratefull to Professors A.H. Heuer and K.P.D. Lagerlof (Case Western Reserve University, Cleveland) for discussions concerning climb dissociation in sapphire. This work has been supported by the Ministry of Science and Technology (Government of Spain) through the project MAT2003-04199-CO2-02.

Figure captions

Figure 1 : Dissociation of a perfect basal dislocation into two partials for (a) sapphire, (b) ruby doped with 725 mol ppm of $Cr^{3+}$ and (c) ruby doped with 9540 mol ppm of $Cr^{3+}$ (the variation of contrast along dislocation C is due to inside-outside contrast due to constrictions and exchange of partials). The specimens were deformed to the lower yield point at about 1300°C.

Figure 2 : Scheme of the electron beam incidence on a climb dissociated dislocation assumed perpendicular to the figure, showing the relation between $\alpha$ and $\beta$ (angles between the stacking fault plane and the normal to the electron beam on the one hand and the [0001] direction on the other hand), $\theta$ (angle between the electron beam and the [0001] direction), $h$ (width of the stacking fault ribbon) and $d$ (separation of partials on the micrograph).

Figure 3 : Micrograph performed at 50º from the basal plane (foil normal to [$22\bar{4}1$]), showing a perfect loop (a) which is climb dissociated, and a typical configuration of faulted and unfaulted loop (b). Sapphire deformed at 1300°C.

Figure 4 : (a) Unfaulted and faulted loops (~~BP~~PL, ~~BF~~FL) correspond to ruby doped with 9540 mol ppm of $Cr^{3+}$, deformed to $\varepsilon = 2\%$ by basal slip at 1300 ºC. One can observe a case of climb dissociation (D) that is magnified 2 times in the insert. The foil is close to [$22\bar{4}1$] axis. PL and FL suggest the propagation by slip of the inner partial dislocation. Similar to dislocation C in Figure 1(c), the contrast of the loop identified as PL is changed from inside to outside in the lower part of its right-hand branch parallel to [$3\bar{3}00$] reflecting the reversal of the relative positions of the partials in the climb-dissociated dislocation. The contrast of this weak beam micrograph is inverted. (b) Schematic view of the mechanism of transformation by glide of a perfect into a faulted dipole.

Figure 5 : Transmission electron micrograph of sapphire deformed by 1.7% at 1250 ºC by basal slip and then maintained 200 min at 804 °C under an applied stress of 451 MPa. A particular configuration of dislocation can be observed that looks like a dissociated superkink labelled SK. Foil normal within 10° from the [$40\bar{4}1$]. (a) stacking fault in contrast g = $01\bar{1}\bar{4}$.

(b) stacking fault out of contrast g = $1\bar{2}10$. Most of the dislocations are heavily cusped as in figures 1b and 1a.

Figure A. The dependence of the repulsive driving force $F_C$ for climb dissociation on the character $\psi$ of the parent dislocation. $F_C^s$ and $F_C^s$ refer to the forces between the screw and edge components of the dislocation partials, respectively. The Burgers vectors of the partials are at 60° from one another.

APPENDIX

In the vast majority of documented cases, such as ordered alloys (e.g. $DO_{22}$ CuZn or $L1_2$ $Ni_3Al$ [26]) and oxides (e.g. spinel and prism plane slip in sapphire [2, 4, 8, 27]), dissociation involves partials with collinear Burgers vectors so that dissociation by climb is precluded in the screw orientation. This is not the case in sapphire where the two partials have Burgers vectors at 60° from one another. In case of a dislocation dissociated according to reaction (1), the climb component of interaction force between the two partials located at a distance $r$ from one another is given by [24]

$$F_C = \frac{\mu b^2}{8\pi r}\left(\frac{2 - \nu - 2\nu\cos(2\psi)}{1 - \nu}\right)$$

where $\psi$ is the angle between the Burgers vector of the parent dislocation and the dislocation line direction. The contributions of the screw and edge components are written

$$F_C^s = \frac{\mu b^2}{8\pi r}\left(1 + 2\cos(2\psi)\right)$$

$$F_C^e = \frac{\mu b^2}{8\pi r}\left(\frac{1 - 2\cos(2\psi)}{1 - \nu}\right)$$

respectively. These formulae are plotted in figure A1.

One can check that the driving force for climb dissociation decreases by less than 10% between the edge and the 60° orientation, and by 30% between edge to the $<1\bar{1}00>$ 30° orientation. As expected the apparent insensitivity of the measured dissociation width to dislocation character (6 to 8 nm between 30° and edge in character) is but moderately influenced by elasticity. Since the fault bordered by a climb-dissociated dislocation changes continuously with line orientation, this result indicates that the energy of the range of stacking fault configurations thus spanned is fairly isotropic.

It is worth noting that the driving force for dissociation, which is of course always positive, results from the contributions of the edge and screw components of the partial dislocations which change sign but in the character ranges where one component is negative, the other is always more positive. It is for $\psi = \dfrac{1}{2}\left( \cos^{-1}\left( \dfrac{\nu}{2(1-\nu)} \right) \right) = 43°$ that the edge and screw components of the partials contribute equally to the repulsive force. For $\psi = 0°$, the screw and edge components of the companion partials are parallel and antiparallel, respectively contributing to the total repulsive force under the ratio of -3(1-$\nu$). In other words, in the screw orientation, the driving force for climb dissociation arises from the repulsive interaction of the screw components in excess of the contribution of the edge components which is attractive. The situation is reversed for $\psi = 90°$.

Table 1. Distance $d$ between the images of companion partial dislocation for various electron beam orientations $\theta$, for sapphire and rubies doped with two different concentrations of chromium ( 725 and 9540 mol ppm of Cr3+) at 1300°C. Examples are shown in Figure 1 corresponding with $\theta$ = 74°, 36° and 54° in sapphire, and ruby doped with 725 and 9540 mol ppm of Cr$^{3+}$, respectively.

| Undoped sapphire | | Sapphire doped with 725 mol ppm of Cr | | Sapphire doped with 9540 mol ppm of Cr | |
|---|---|---|---|---|---|
| $\theta$ (°) | $d$ (nm) | $\theta$ (°) | $d$ (nm) | $\theta$ (°) | $d$ (nm) |
| 38 ± 4 | 4.0 – 6.8 | 36 ± 4 | 2.7 - 7.1 | 44 ± 4 | 5.5 - 9.9 |
| 44 ± 4 | 3.1 – 6.5 | 48 ± 4 | 4.3 - 7.3 | 54 ± 4 | 6.7 - 10.9 |
| 49 ± 4 | 4.2 – 7.0 | 58 ± 4 | 5.9 - 9.7 | 61 ± 4 | 7.5 - 11.5 |
| 64 ± 4 | 4.2 – 7.6 | 67 ± 4 | 5.9 -10.1 | 69 ± 4 | 7.3 - 10.9 |
| 74 ± 4 | 5.1 – 8.9 | 77 ± 4 | 6.2 -10 | 80 ± 4 | 4.9 - 8.3 |
| 85 ± 4 | 4.8 – 8.2 | 98 ± 4 | 5.6 -10.4 | 85 ± 4 | 7.8 - 10.8 |
| 95 ± 4 | 4.9 – 8.5 | | | | |
| 102 ± 4 | 5.4 – 7.8 | | | | |

Table 2: $\beta$ and $h$ values providing the best fit to expression 3.4.2 for undoped and doped sapphire. It shows that chromium does not affect the stacking fault plane, since dissociation occurs by nearly pure climb. A decrease of the stacking fault energy due to an increase of the distance $h$ between partial dislocations are found.

| Sample | $\beta$ (°) | $h$ (nm) |
|---|---|---|
| Undoped sapphire | 3 - 12 | 5 - 8 |
| Sapphire doped with 725 mol ppm of $Cr^{3+}$ | 18 - 20 | 5 - 8 |
| Sapphire doped with 9540 mol ppm of $Cr^{3+}$ | 11 - 16 | 7 - 10 |

**Figure 1a**
Click here to download high resolution image

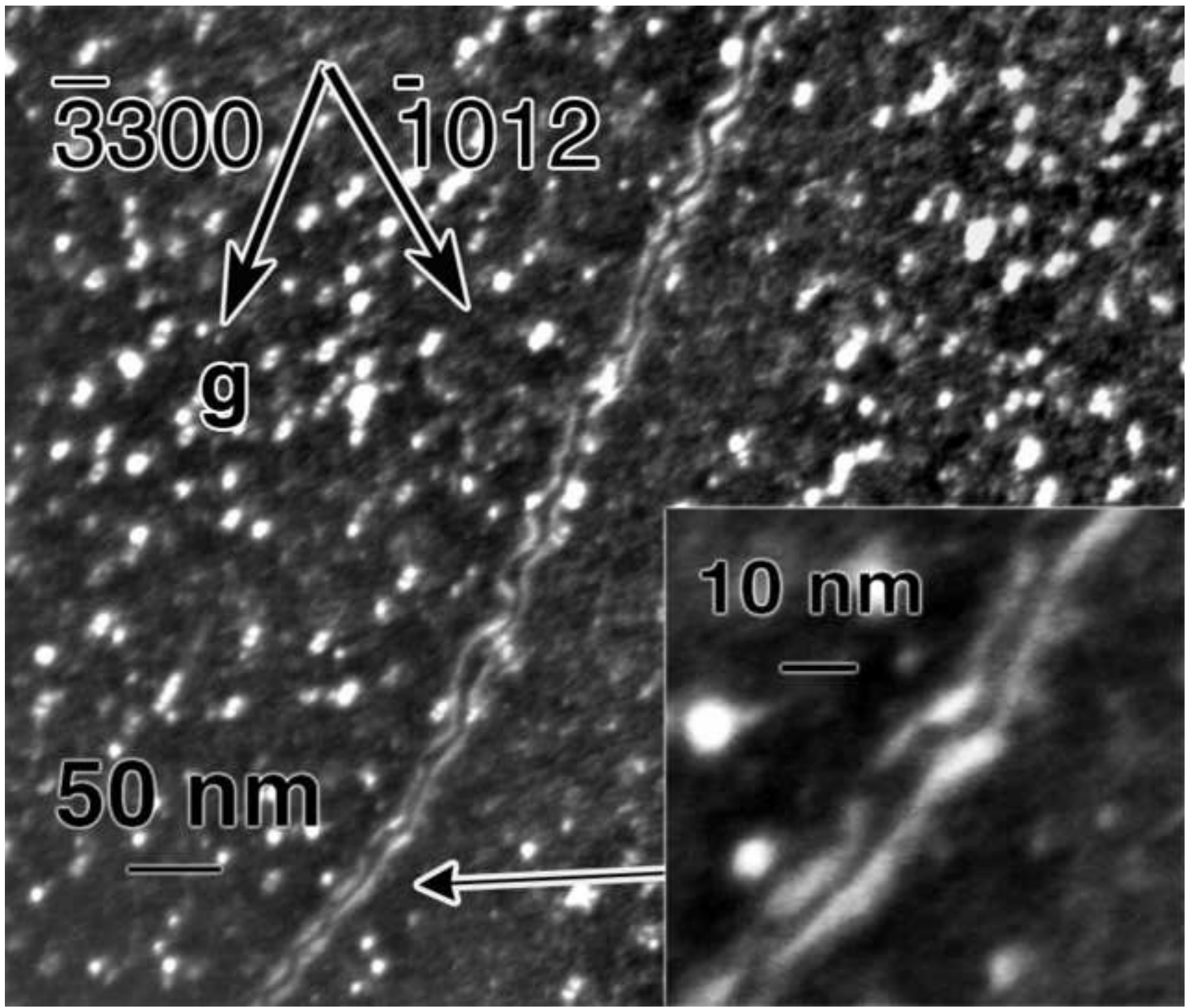

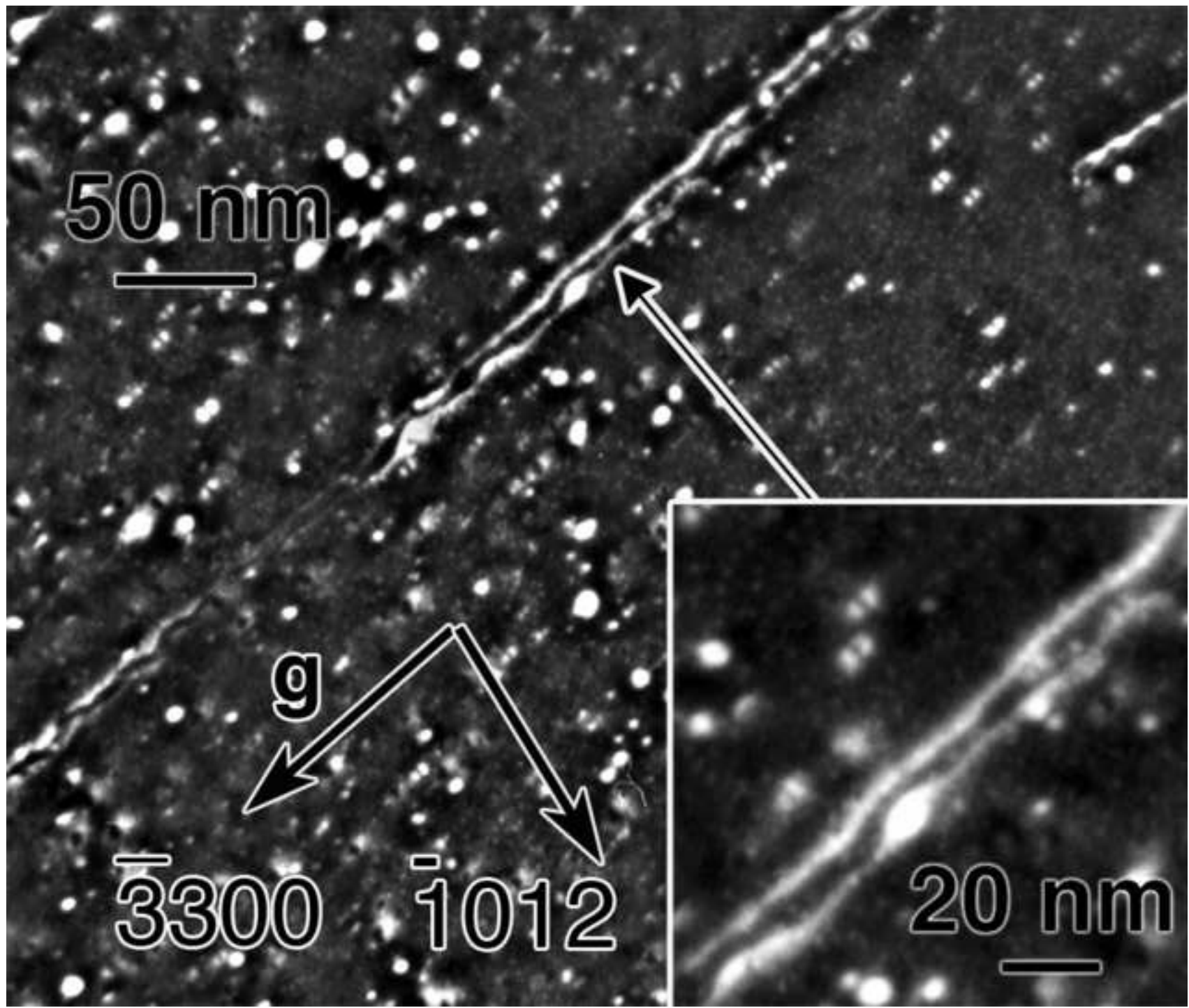

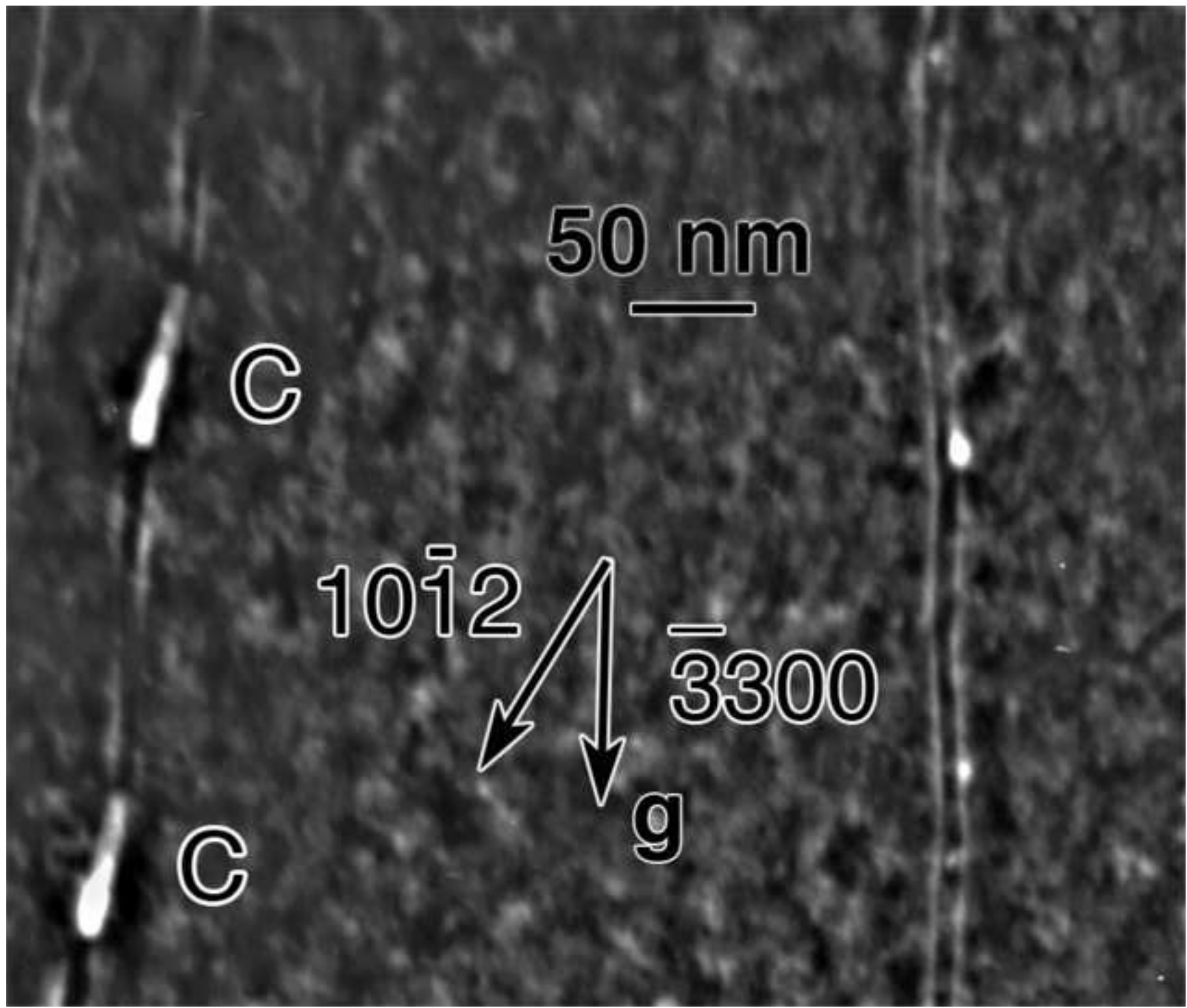

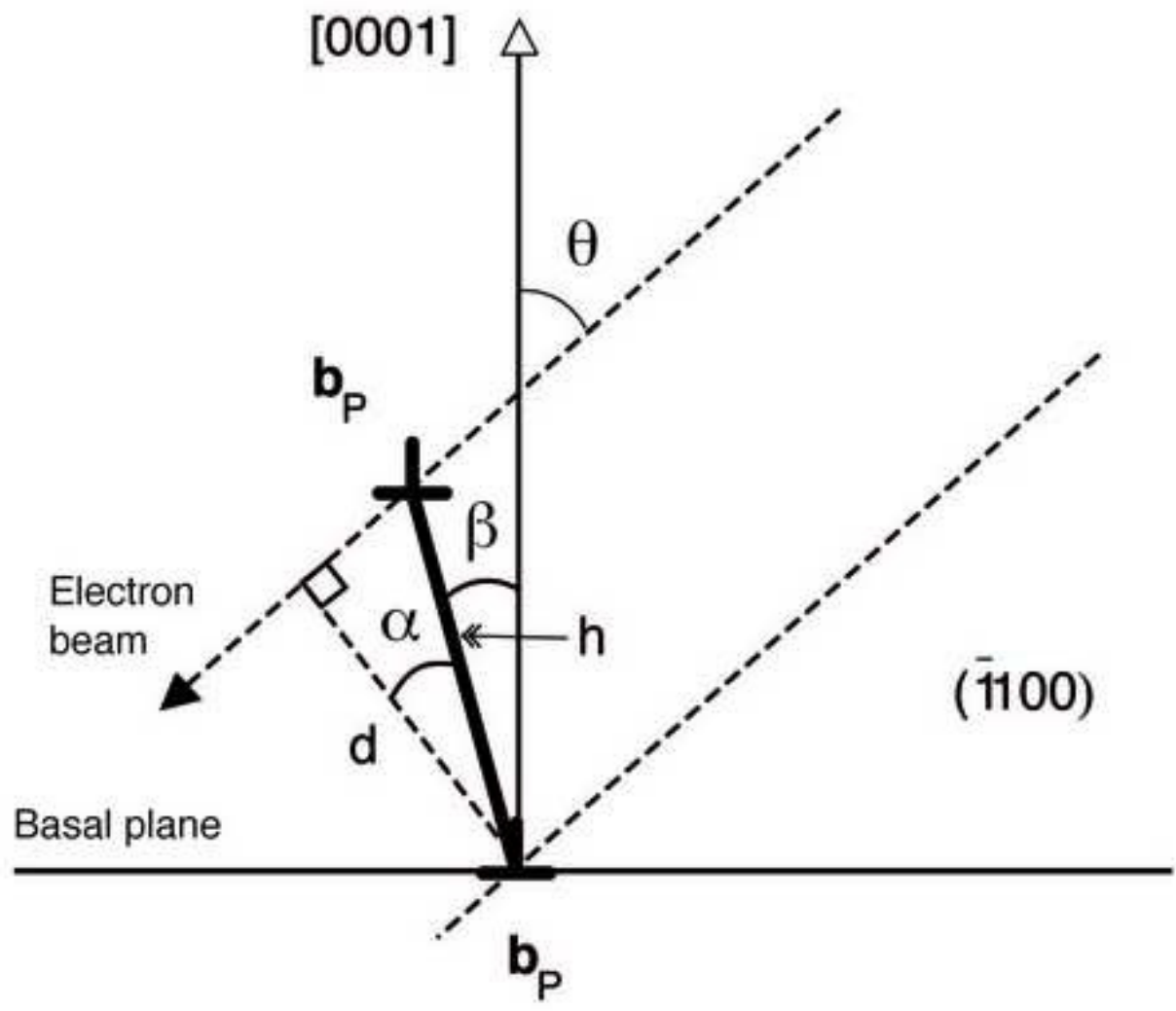

**Figure 3**

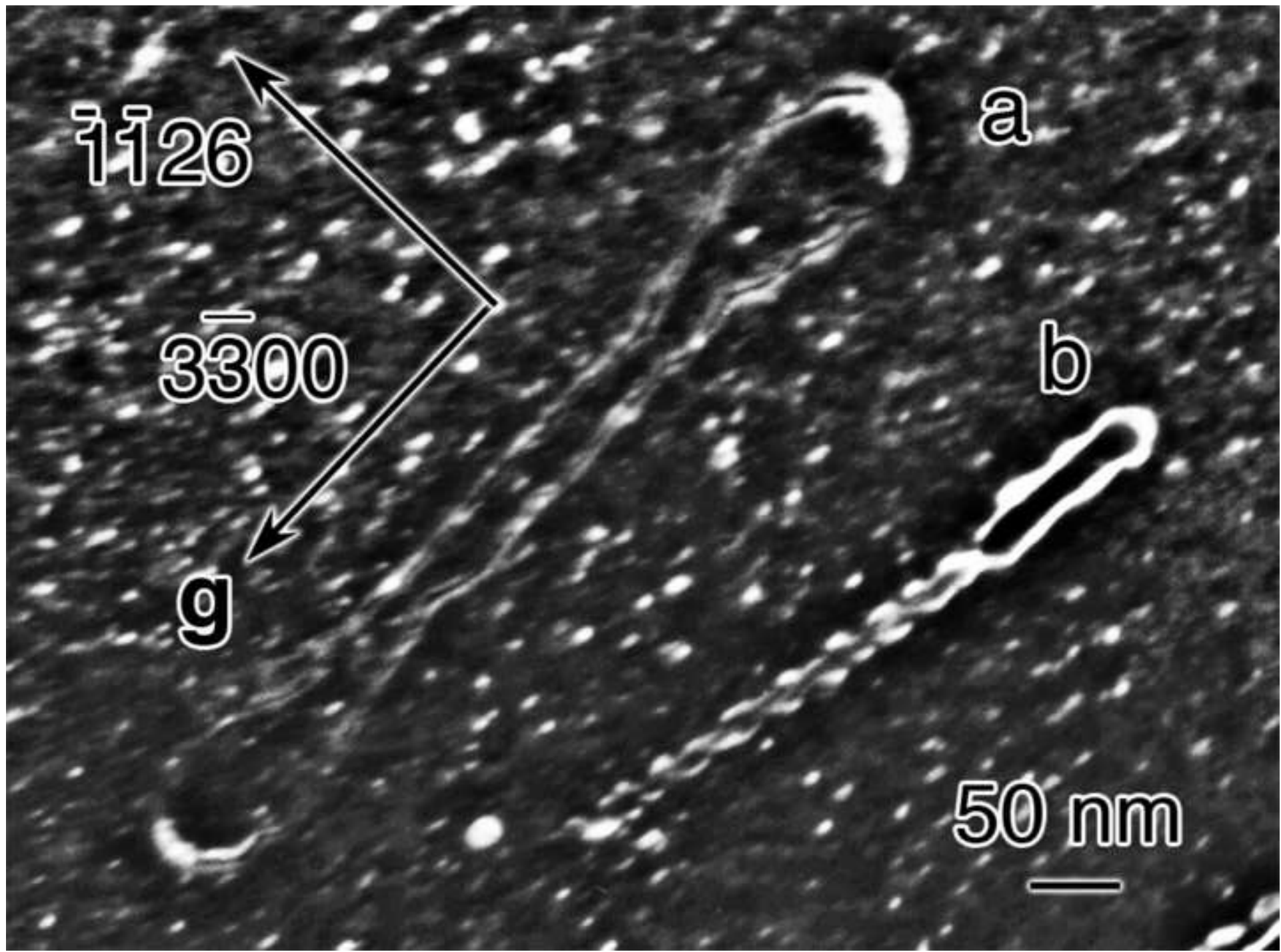

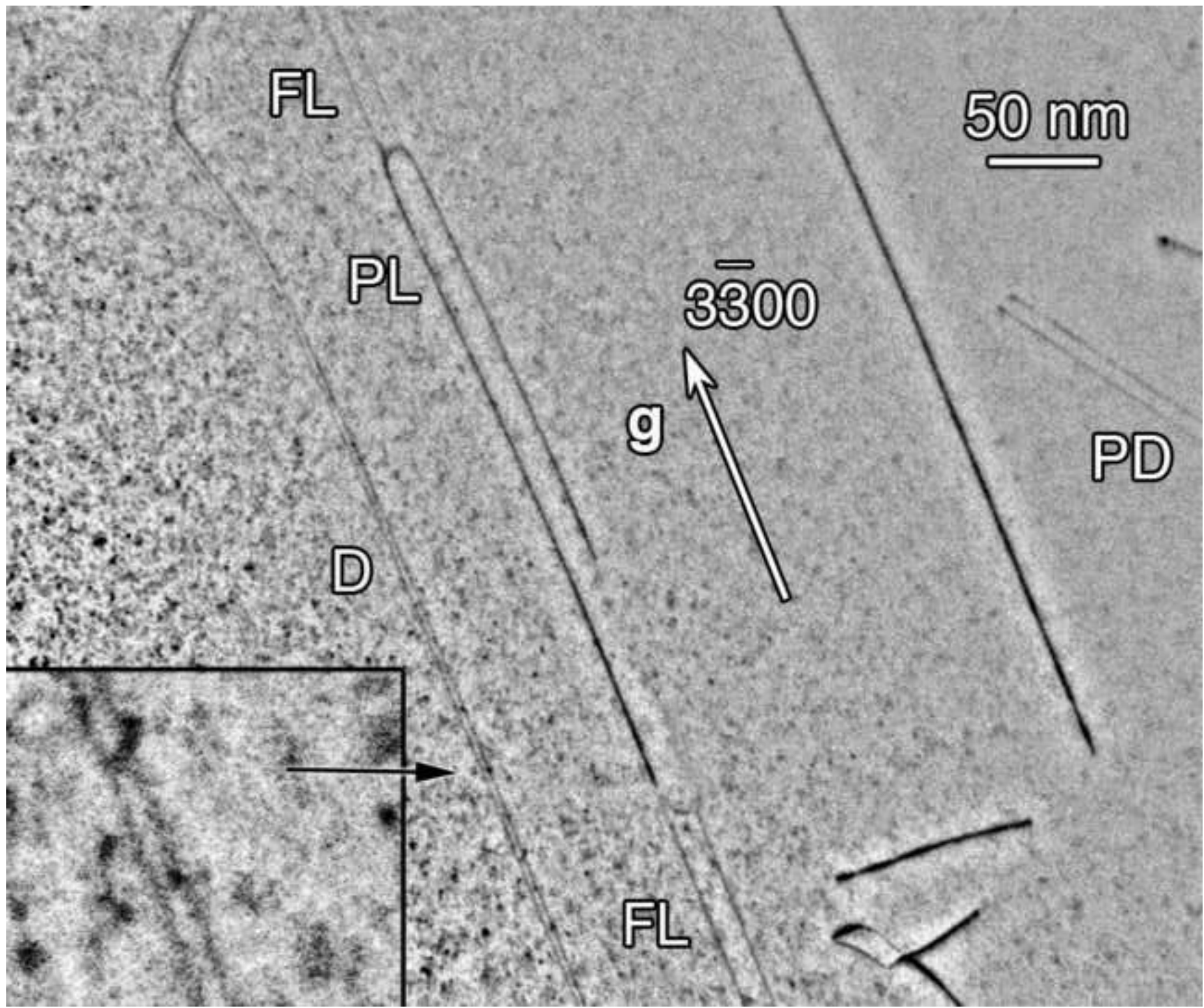

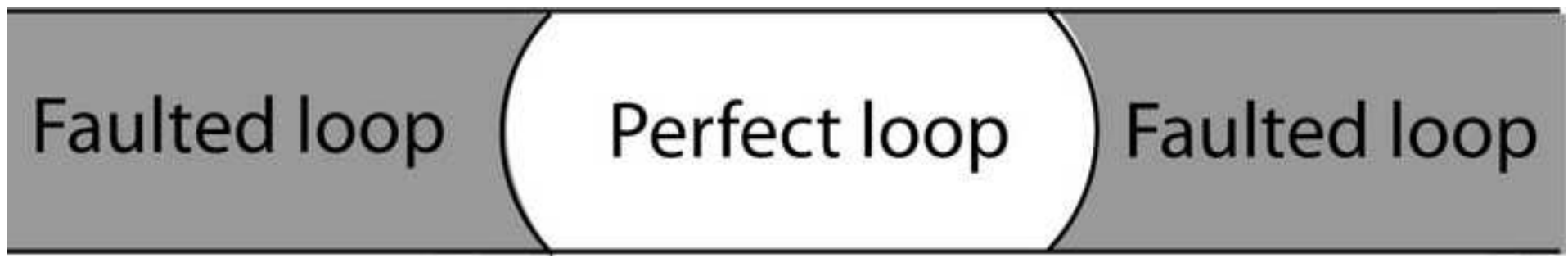

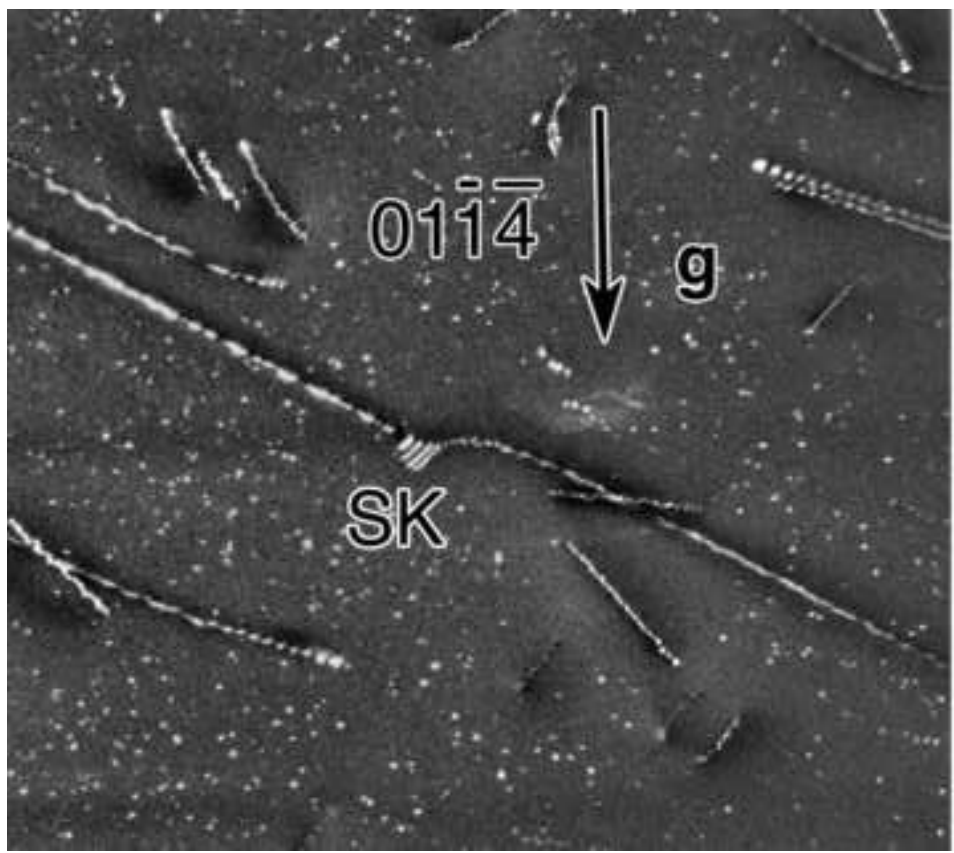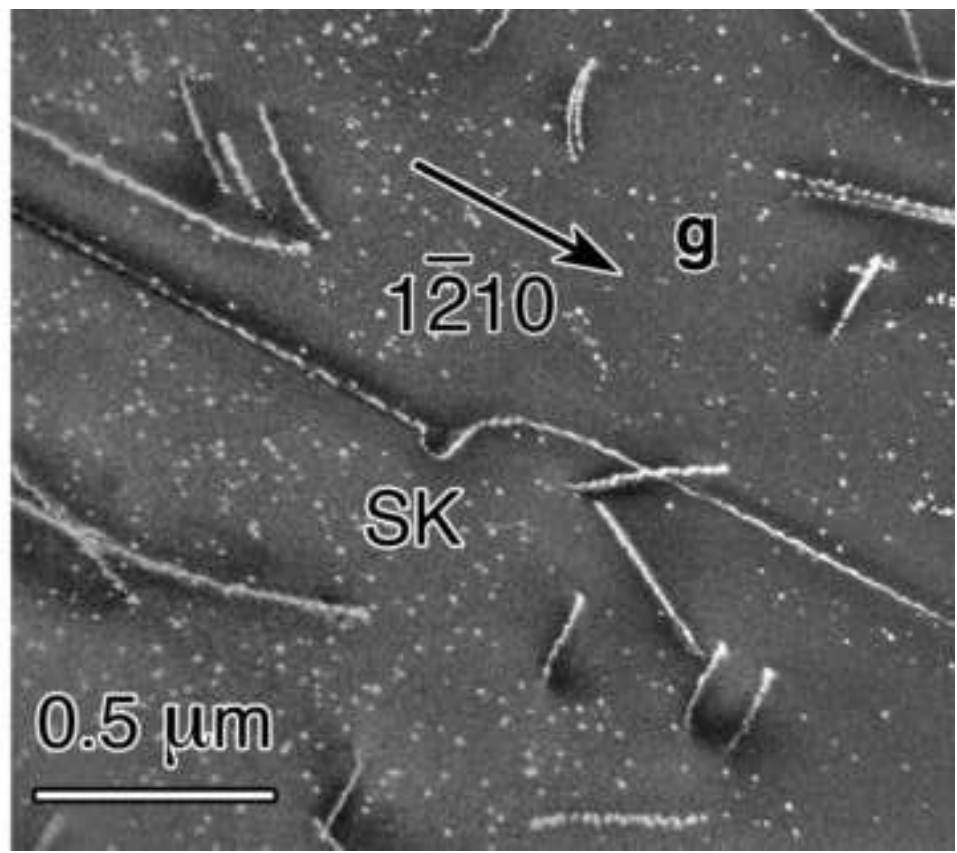

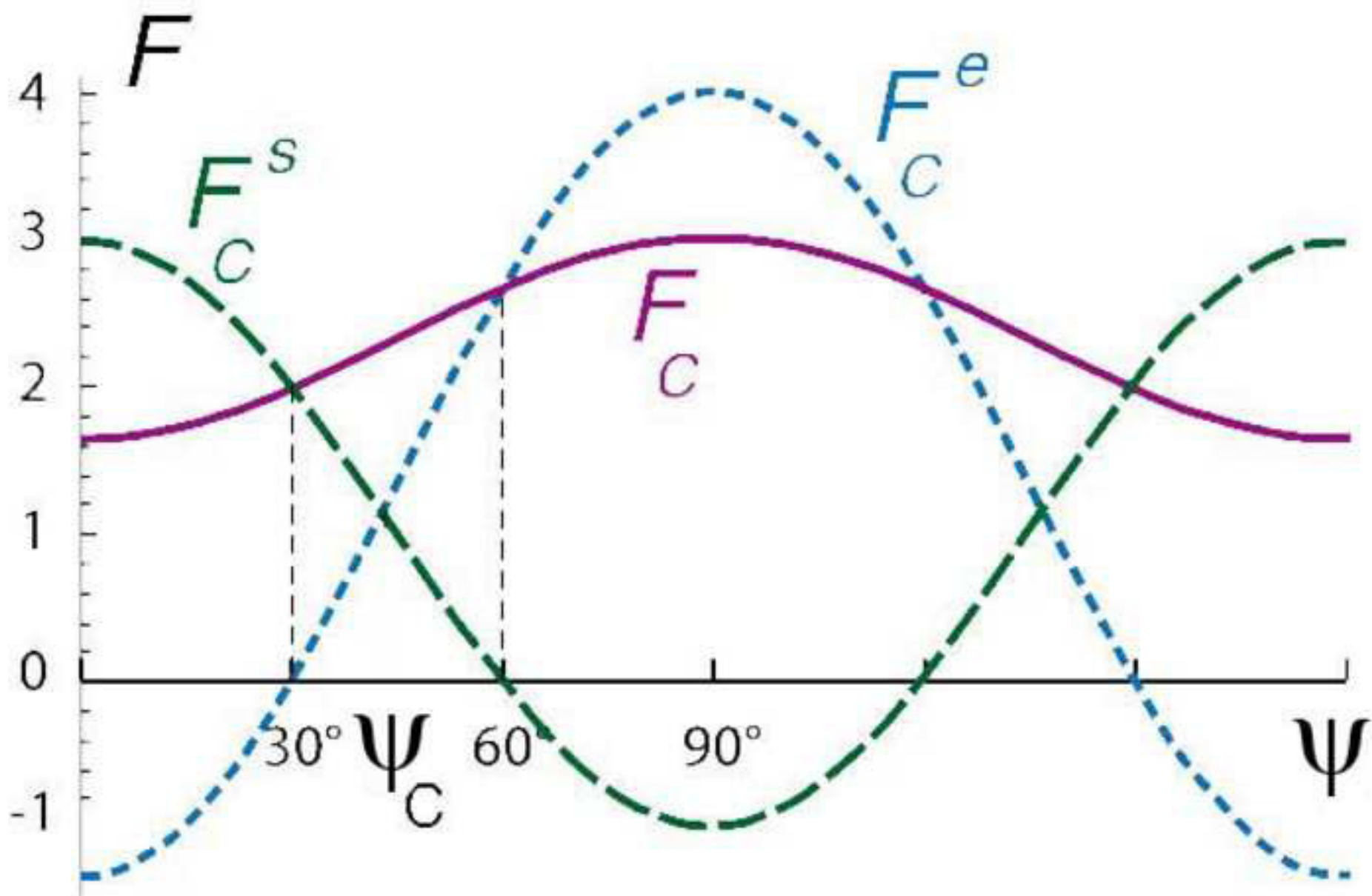